\documentclass[superscriptaddress,amsmath,amssymb,aps,pra]{revtex4-2}

\DeclareUnicodeCharacter{2032}{\ensuremath{\prime}}

\usepackage{graphicx}
\usepackage{dcolumn}
\usepackage{bm}
\usepackage[colorlinks]{hyperref}
\usepackage{float}
\usepackage{appendix}
\usepackage{physics}
\usepackage{epigraph}

\def\beq{\begin{equation}}
\def\eeq{\end{equation}}
\def\bsp{\begin{split}}
\def\esp{\end{split}}
\def\bea{\begin{eqnarray}}
\def\eea{\end{eqnarray}}

\begin{document}

\title{The time qubit}

\author{Yakov Bloch}
\affiliation{Raymond and Beverly Sackler School of Physics and Astronomy, Tel-Aviv University, IL-69978 Tel Aviv, Israel}

\begin{abstract}
A spin precessing in a magnetic field is often used as a quantum clock, for example in tunneling-time measurements. We show that such a clock can exist in a coherent superposition of opposite temporal orientations, treating the arrow of time as a quantum two-level system. A Mach-Zehnder interferometer with equal and opposite magnetic fields provides a simple implementation, enabling interference between forward and backward evolution and Bell-type tests of temporal coherence. Extending the framework to relativistic dynamics reveals that the Dirac Hamiltonian describes an intrinsic coupling between a spin qubit and a time qubit, with matter and antimatter corresponding to opposite poles on the temporal Bloch sphere.
\end{abstract}

\maketitle

\section{Introduction}

A spin-$\tfrac{1}{2}$ particle precessing in a magnetic field provides one of the most elementary realizations of a quantum clock: its angular phase advances at a constant rate and serves as an operational measure of elapsed time \cite{PhysRevB.27.6178, RevModPhys.61.917, RevModPhys.66.217, PhysRevLett.127.133001, PhysRevA.53.3749,bm9q-r6j9, PhysRevA.53.434, PhysRevA.59.2261, PhysRevA.105.062804, PhysRevA.64.042112, PhysRevB.38.3287, PhysRevB.40.5387, martin1992theory, sokolovski2021tunnelling}. Spin-based clocks have long been central to studies of tunneling and dwell times, where the accumulated spin phase records how long a particle interacts with a potential barrier. Because the precession is directly governed by the spin Hamiltonian, this system offers a direct link between quantum dynamics and temporal measurement.

At the same time, this simplicity exposes a fundamental question. In standard quantum mechanics, time enters as an external parameter rather than as an operator, and the arrow of time, the distinction between forward and backward evolution, is imposed through boundary conditions rather than determined dynamically. If a spin’s precession functions as a clock, can the direction of its evolution itself become a quantum variable? Reversing the magnetic field reverses the precession, corresponding to a change in the sign of the Hamiltonian. The contrast between clockwise and counter-clockwise rotation thus mirrors a reversal of temporal orientation. The notions of ``forward'' and ``backward'' evolution, usually treated as fixed background structure, can in principle be coherently controlled and superposed.

We show that this control can be realized in an experimentally accessible system. A spin-$\tfrac{1}{2}$ particle traversing a Mach-Zehnder interferometer with magnetic fields of equal magnitude and opposite direction in its two arms simultaneously experiences both orientations of the field. In one arm the spin precesses clockwise, in the other counter-clockwise, producing a coherent superposition of two temporal orientations. The interferometer’s path degree of freedom thereby acts as a two-level system labeling the sign of the time-evolution generator, a time qubit whose logical basis states correspond to opposite arrows of time.

When the paths recombine, the amplitudes associated with these opposite time directions interfere. The two output ports correspond to symmetric and antisymmetric superpositions of the path states and therefore project onto components of the spin dynamics that are even and odd under time reversal. Detection at one port isolates the time-symmetric contribution, while detection at the other isolates the antisymmetric one. The resulting interference pattern represents a set of time-parity fringes, a direct manifestation of quantum interference between forward- and backward-evolving histories of a single clock.

The proposed setup is, in essence, an interferometer in temporal orientation. The two arms represent opposite directions of time flow, and the second beam splitter performs a measurement in the time-parity basis. The path qubit, normally a spatial variable, thus acts as a time-orientation qubit whose coherent superpositions embody an indefinite arrow of time. Such states can be prepared, manipulated, and measured using the same techniques developed for conventional qubits, offering an operational framework for describing coherent superpositions of temporal orientations.

When the time qubit is entangled with the spin degree of freedom, the composite system enables Bell-type tests \cite{bell2004speakable, greenberger1989going, greenberger1990bell} in which measurements of noncommuting time qubit observables and different spin components produce correlations that violate a CHSH inequality \cite{khrennikov2015chsh, PhysRevLett.88.060403, PhysRevA.88.052105, PhysRevA.73.022110}. In this way, temporal orientation may turn out to behave as a genuine quantum observable, capable of superposition and entanglement on the same footing as spin.

Conceptually, this unifies several lines of thought that have remained distinct. In the two-state-vector pre- and post-selection formalism \cite{hashmi2016two, aharonov2014measurement, aharonov2008two, PhysRevA.98.032312, PhysRevA.41.11}, a system is described simultaneously by forward- and backward-evolving states, but these are typically treated as boundary conditions for a single process rather than as an explicit, controllable degree of freedom. In theories of indefinite causal order \cite{PhysRevA.105.062216, PhysRevLett.121.090503, rubino2017experimental, rozema2024experimental}, temporal orderings of operations are coherently superposed, though usually at an abstract level. In relational \cite{rovelli1996relational, calosi2024relational, ruyant2018can, martin2019notion} or Page-Wootters \cite{PhysRevD.27.2885, qfns-48vq, PhysRevA.103.052420, PhysRevResearch.4.013180, PhysRevD.108.063507} models, time emerges from correlations between system and clock, but the arrow of time is fixed by construction. The present approach integrates these perspectives within a single, realizable experiment, demonstrating that forward and backward evolutions can coexist, interfere, and generate measurable nonclassical correlations.

Extending the framework to relativistic dynamics reveals that the same two-level temporal structure underlies the Dirac equation. When expressed in time qubit form, the Dirac Hamiltonian describes a spin qubit coupled to a temporal qubit, with $\tau_z$ representing the sign of energy, distinguishing matter from antimatter, and $\tau_x$ generating coherent mixing between opposite temporal orientations through spatial motion. The familiar zitterbewegung corresponds to precession of the temporal Bloch vector, while pair creation and annihilation can be viewed as transitions between its two orientations. In this picture, mass acts as a longitudinal field that stabilizes a definite temporal direction, and the Majorana case corresponds to a fixed superposition of forward and backward components. The time-qubit formulation therefore unifies interferometric, nonrelativistic, and relativistic aspects of temporal symmetry within a single Hamiltonian framework.

Beyond its conceptual significance, the proposal establishes an operational framework that promotes temporal orientation to a fully quantum degree of freedom. The introduction of the time qubit provides a language for describing and quantifying coherence, entanglement, and contextuality in time direction itself. Potential applications range from time-based interferometric metrology, where opposite evolutions enhance sensitivity to Hamiltonian asymmetries, to foundations of quantum thermodynamics and causality, where temporal coherence may play a role analogous to spatial entanglement. In this sense, the time qubit formalism offers both an experimentally realizable scenario and a theoretical paradigm in which the arrow of time becomes an active quantum resource.

\section{The time qubit}

To formalize the notion of a time qubit, we consider a composite Hilbert space
\begin{equation}
\mathcal{H} = \mathcal{H}_{\mathrm{S}} \otimes \mathcal{H}_{\mathrm{T}},
\label{hilbert}
\end{equation}
where $\mathcal{H}_{\mathrm{S}}$ describes the physical system and $\mathcal{H}_{\mathrm{T}}$ is a two-dimensional Hilbert space encoding temporal orientation. The logical basis of the time qubit, \(\{|+\rangle_{\mathrm{T}}, |-\rangle_{\mathrm{T}}\}\), represents evolution with opposite signs of the time-evolution generator.

We associate to the two temporal orientations a pair of Hamiltonians
\(\hat{H}_{+}\) and \(\hat{H}_{-}\) that differ by the sign of a term
\(\hat{V}\) responsible for the direction of dynamical evolution:
\begin{equation}
    \hat{H}_{\pm} = \hat{H}_0 \pm \hat{V},
    \label{eq:Hpm}
\end{equation}
where $\hat{H}_0$ is the time-symmetric part of the dynamics.
The composite or ``controlled'' Hamiltonian on
$\mathcal{H}_{\mathrm{S}}\!\otimes\!\mathcal{H}_{\mathrm{T}}$ is then
\begin{equation}
    \hat{H}_{\mathrm{tot}}
      = |+\rangle\!\langle+|_{\mathrm{T}}\!\otimes\!\hat{H}_{+}
       + |-\rangle\!\langle-|_{\mathrm{T}}\!\otimes\!\hat{H}_{-}.
    \label{eq:HtotProjectors}
\end{equation}
Writing the projectors in terms of the Pauli operator
\(\tau_z = |+\rangle\!\langle+|_{\mathrm{T}}
            - |-\rangle\!\langle-|_{\mathrm{T}}\),
we obtain the compact form
\begin{equation}
    \hat{H}_{\mathrm{tot}}
      = \mathbb{I}_{\mathrm{T}}\!\otimes\!\hat{H}_0
        + \tau_z\!\otimes\!\hat{V}.
    \label{eq:HtotCompact}
\end{equation}
The state of the time qubit, being an eigenstate of $\tau_z$ (with eigenvalues \(\pm1\)) selects whether the system evolves with
$\hat{H}_+$ or $\hat{H}_-$.

As an example, we consider the Zeeman Hamiltonian (other implementations of a time qubit may include driven two-level systems with sign-reversed detuning, optical interferometers with reversed phase accumulation and dirac-like or relativistic analog Systems). For a spin-$\tfrac{1}{2}$ particle the Zeeman Hamiltonian in a uniform magnetic
field $\mathbf{B}$ is
\begin{equation}
    \hat{H}_{+}
      = \frac{\hat{p}^2}{2m}
        - \frac{g\mu_B}{2}\,\boldsymbol{\sigma}\!\cdot\!\mathbf{B}.
    \label{eq:HplusZeeman}
\end{equation}
We define the opposite orientation of evolution by reversing the direction of
the field,
\begin{equation}
    \hat{H}_{-}
      = \frac{\hat{p}^2}{2m}
        - \frac{g\mu_B}{2}\,\boldsymbol{\sigma}\!\cdot\!(-\mathbf{B})
      = \frac{\hat{p}^2}{2m}
        + \frac{g\mu_B}{2}\,\boldsymbol{\sigma}\!\cdot\!\mathbf{B}.
    \label{eq:HminusZeeman}
\end{equation}
The two Hamiltonians generate Larmor precessions with opposite directions. 

The corresponding time-evolution operators are
\begin{equation}
    \hat{U}_{\pm}(t)
      = e^{-\tfrac{i}{\hbar}\hat{H}_{\pm}t}.
\end{equation}
The general state of the composite system can be written as
\begin{equation}
|\Psi(0)\rangle =
\alpha|+\rangle_{\mathrm{T}}\!\otimes|\psi_+^{(0)}\rangle
+\beta|-\rangle_{\mathrm{T}}\!\otimes|\psi_-^{(0)}\rangle,
\label{eq:initState}
\end{equation}
which evolves to
\begin{equation}
|\Psi(t)\rangle =
\alpha|+\rangle_{\mathrm{T}}\!\otimes
 e^{-i\hat{H}_+t/\hbar}|\psi_+^{(0)}\rangle
+\beta|-\rangle_{\mathrm{T}}\!\otimes
 e^{-i\hat{H}_-t/\hbar}|\psi_-^{(0)}\rangle.
\end{equation}
Each temporal branch follows its corresponding Hamiltonian, and measurements on
the time qubit select definite or coherent combinations of the two directions.

In the basis
\[
|\!\pm_x\rangle_{\mathrm{T}}
  = \frac{1}{\sqrt{2}}\bigl(|+\rangle_{\mathrm{T}}
                            \pm|-\rangle_{\mathrm{T}}\bigr),
\]
projection of the time qubit onto $|\!\pm_x\rangle_{\mathrm{T}}$ yields
the time-even and time-odd components of the system’s evolution:
\begin{align}
    \hat{U}_{\mathrm{even}}(t)
      &= \tfrac{1}{2}\!\left(
        e^{-\frac{i}{\hbar}\hat{H}_{+}t}
      + e^{-\frac{i}{\hbar}\hat{H}_{-}t}\right),\\[4pt]
    \hat{U}_{\mathrm{odd}}(t)
      &= \tfrac{1}{2}\!\left(
        e^{-\frac{i}{\hbar}\hat{H}_{+}t}
      - e^{-\frac{i}{\hbar}\hat{H}_{-}t}\right).
    \label{eq:EvenOdd}
\end{align}
These are the conditional Kraus operators acting on the system when the time
qubit is prepared and measured in the $\tau_x$ basis.
They correspond respectively to processes that are symmetric and antisymmetric
under exchange of the time direction, analogous to even and odd parity under
spatial reflection.

The introduction of the time qubit elevates the discrete symmetry of time
reversal into an observable quantum degree of freedom.
The total Hilbert space supports coherent superpositions of forward and
backward temporal orientations, while measurements in complementary time-qubit
bases access different linear combinations of advanced and retarded
evolutions.
In this way, the arrow of time becomes a controllable quantum variable whose
state can be manipulated, entangled with other observables, and measured within
a single unified framework.

\section{Bloch-sphere representation of the time qubit}
\label{sec:BlochTimeQubit}

The time qubit is a two-level system that encodes the sign of the generator of
time evolution. Its Hilbert space $\mathcal{H}_{\mathrm{T}}$ is spanned by the
orthonormal basis
\begin{equation}
    \{|+\rangle_{\mathrm{T}},|-\rangle_{\mathrm{T}}\},
\end{equation}
where $|+\rangle_{\mathrm{T}}$ corresponds to evolution with a Hamiltonian
$\hat{H}_{+}$ and $|-\rangle_{\mathrm{T}}$ to evolution with the time-reversed
Hamiltonian $\hat{H}_{-}$. On $\mathcal{H}_{\mathrm{T}}$ we define the Pauli
operators
\begin{equation}
    \tau_z
      = |+\rangle_{\mathrm{T}}\!\langle+|
       -|-\rangle_{\mathrm{T}}\!\langle-|, \qquad
    \tau_x
      = |+\rangle_{\mathrm{T}}\!\langle-|
       +|-\rangle_{\mathrm{T}}\!\langle+|, \qquad
    \tau_y
      = -i|+\rangle_{\mathrm{T}}\!\langle-|
       +i|-\rangle_{\mathrm{T}}\!\langle+|.
\end{equation}
The operator $\tau_z$ distinguishes definite temporal orientation (forward
vs.~backward), while $\tau_x$ and $\tau_y$ probe coherent superpositions of
these orientations.

Any pure state of the time qubit can be written as a superposition
\begin{equation}
    |\psi_{\mathrm{T}}\rangle
      = \alpha |+\rangle_{\mathrm{T}}
       + \beta  |-\rangle_{\mathrm{T}},
    \qquad
    |\alpha|^2+|\beta|^2=1.
\end{equation}
Introducing polar angles $(\theta,\varphi)$ on the Bloch sphere, we may write
this state in the standard parametrization
\begin{equation}
    |\psi_{\mathrm{T}}(\theta,\varphi)\rangle
      = \cos\!\left(\frac{\theta}{2}\right)|+\rangle_{\mathrm{T}}
        + e^{i\varphi}\sin\!\left(\frac{\theta}{2}\right)|-\rangle_{\mathrm{T}},
    \qquad
    0\le\theta\le\pi,\;\;0\le\varphi<2\pi.
    \label{eq:TimeQubitBlochState}
\end{equation}
The corresponding Bloch vector
$\boldsymbol{n} = (n_x,n_y,n_z)$ is defined by the expectation values of the
Pauli operators,
\begin{equation}
    n_k
      = \langle\psi_{\mathrm{T}}|\tau_k|\psi_{\mathrm{T}}\rangle,
    \qquad
    k\in\{x,y,z\}.
\end{equation}
For the state \eqref{eq:TimeQubitBlochState} one finds
\begin{equation}
    n_x
      = \sin\theta\cos\varphi, \qquad
    n_y
      = \sin\theta\sin\varphi, \qquad
    n_z
      = \cos\theta.
\end{equation}
Thus any pure time qubit state corresponds to a unit vector
$\boldsymbol{n}\in\mathbb{R}^3$ with $|\boldsymbol{n}|=1$ on the Bloch sphere.
The ``north pole'' $\boldsymbol{n}=(0,0,1)$ represents the forward-evolving
state $|+\rangle_{\mathrm{T}}$, while the ``south pole''
$\boldsymbol{n}=(0,0,-1)$ represents the backward-evolving state
$|-\rangle_{\mathrm{T}}$.

The eigenstates of $\tau_z$,
\begin{equation}
    \tau_z|+\rangle_{\mathrm{T}} = +|+\rangle_{\mathrm{T}}, \qquad
    \tau_z|-\rangle_{\mathrm{T}} = -|-\rangle_{\mathrm{T}},
\end{equation}
represent definite temporal orientations: in our construction they label
whether the system evolves under $\hat{H}_{+}$ or under $\hat{H}_{-}$. In the
Bloch-sphere picture, they are located at the two poles:
\begin{equation}
    |+\rangle_{\mathrm{T}}
      \;\leftrightarrow\; (n_x,n_y,n_z) = (0,0,1),\qquad
    |-\rangle_{\mathrm{T}}
      \;\leftrightarrow\; (0,0,-1).
\end{equation}

The eigenstates of $\tau_x$,
\begin{equation}
    |\!+\!_x\rangle_{\mathrm{T}}
      = \frac{1}{\sqrt{2}}\bigl(|+\rangle_{\mathrm{T}}
                               +|-\rangle_{\mathrm{T}}\bigr),\qquad
    |\!-\!_x\rangle_{\mathrm{T}}
      = \frac{1}{\sqrt{2}}\bigl(|+\rangle_{\mathrm{T}}
                               -|-\rangle_{\mathrm{T}}\bigr),
\end{equation}
correspond to coherent superpositions of forward and backward evolution. On
the Bloch sphere they lie on the equator along the $x$ axis:
\begin{equation}
    |\!+\!_x\rangle_{\mathrm{T}}
      \;\leftrightarrow\; (1,0,0),\qquad
    |\!-\!_x\rangle_{\mathrm{T}}
      \;\leftrightarrow\; (-1,0,0).
\end{equation}
As we shall see, in the Mach-Zehnder realization these states correspond to the symmetric and
antisymmetric superpositions of the two arms and are associated with the two
output ports, $D_1$ and $D_2$. Measurement in the $\tau_z$ basis probes definite time direction (which Hamiltonian acts), while measurement in
the $\tau_x$ basis probes time parity, that is, the coherence between
forward and backward evolution.

Similarly, the eigenstates of $\tau_y$,
\begin{equation}
    |\!+\!_y\rangle_{\mathrm{T}}
      = \frac{1}{\sqrt{2}}\bigl(|+\rangle_{\mathrm{T}}
                               +i|-\rangle_{\mathrm{T}}\bigr),\qquad
    |\!-\!_y\rangle_{\mathrm{T}}
      = \frac{1}{\sqrt{2}}\bigl(|+\rangle_{\mathrm{T}}
                               -i|-\rangle_{\mathrm{T}}\bigr),
\end{equation}
lie on the equator along the $y$ axis:
\begin{equation}
    |\!+\!_y\rangle_{\mathrm{T}}
      \;\leftrightarrow\; (0,1,0),\qquad
    |\!-\!_y\rangle_{\mathrm{T}}
      \;\leftrightarrow\; (0,-1,0).
\end{equation}
These states represent superpositions of temporal orientations with a $\pi/2$
relative phase and can be used to probe complementary aspects of time
coherence.

In general, the time qubit will be entangled with the physical system
$\mathcal{H}_{\mathrm{S}}$ and may be in a mixed state. The reduced density
matrix of the time qubit is obtained by tracing over $\mathcal{H}_{\mathrm{S}}$,
\begin{equation}
    \rho_{\mathrm{T}}
      = \operatorname{Tr}_{\mathrm{S}}\!\bigl[\rho_{\mathrm{TS}}\bigr]
      = \frac{1}{2}\Bigl(
           \mathbb{I}_{\mathrm{T}}
          +\boldsymbol{r}\!\cdot\!\boldsymbol{\tau}
        \Bigr),
    \qquad
    \boldsymbol{r}=(r_x,r_y,r_z)\in\mathbb{R}^3,\;\;|\boldsymbol{r}|\le 1,
    \label{eq:TimeQubitMixed}
\end{equation}
where $\boldsymbol{\tau}=(\tau_x,\tau_y,\tau_z)$ and
$\boldsymbol{r}$ is the Bloch vector of the mixed state. The components are
\begin{equation}
    r_k = \operatorname{Tr}\bigl(\rho_{\mathrm{T}}\tau_k\bigr),\qquad
    k\in\{x,y,z\}.
\end{equation}
Pure states correspond to $|\boldsymbol{r}|=1$, while $|\boldsymbol{r}|<1$
signals partial decoherence between time orientations. In particular, $r_z$ quantifies the imbalance between forward and backward evolution probabilities, that is, the bias of the arrow of time, while $r_x$ and $r_y$ quantify the coherence between forward and backward time directions. The length $\sqrt{r_x^2+r_y^2}$ measures the visibility of time-parity fringes. In the Mach-Zehnder realization, environmental decoherence or which-path
information reduces $|\boldsymbol{r}|$ by suppressing $r_x$ and $r_y$ while
leaving $r_z$ approximately unchanged. In the Bloch-sphere picture, loss of coherence between forward and backward time directions suppresses the equatorial components $r_x$ and $r_y$, while the population imbalance $r_z$ remains essentially unchanged. Geometrically, the Bloch vector, which initially lies on the surface of the sphere, then contracts in length and moves closer to the $z$ axis, representing a mixed state with reduced temporal coherence but a definite bias toward one time orientation.

When the time qubit is subjected to a Hamiltonian acting solely on
$\mathcal{H}_{\mathrm{T}}$,
\begin{equation}
    \hat{H}_{\mathrm{T}}
      = \frac{\hbar\Omega_{\mathrm{T}}}{2}\,
        \boldsymbol{n}_0\!\cdot\!\boldsymbol{\tau},
    \qquad
    |\boldsymbol{n}_0|=1,
\end{equation}
its evolution is a rigid rotation of the Bloch vector $\boldsymbol{r}(t)$
around the axis $\boldsymbol{n}_0$ with angular frequency
$\Omega_{\mathrm{T}}$. The corresponding unitary is
\begin{equation}
    \hat{U}_{\mathrm{T}}(t)
      = \exp\!\left(
          -\frac{i}{\hbar}\hat{H}_{\mathrm{T}}t
        \right)
      = \exp\!\left(
          -\frac{i}{2}\Omega_{\mathrm{T}}t\,
          \boldsymbol{n}_0\!\cdot\!\boldsymbol{\tau}
        \right).
\end{equation}
In the full time qubit construction, the time qubit couples to the physical
system via a controlled Hamiltonian of the form
\begin{equation}
    \hat{H}_{\mathrm{tot}}
      = \mathbb{I}_{\mathrm{T}}\otimes\hat{H}_0
        + \tau_z\otimes\hat{V},
\end{equation}
so that $\tau_z$ selects whether the system evolves with
$\hat{H}_{+}=\hat{H}_0+\hat{V}$ or with $\hat{H}_{-}=\hat{H}_0-\hat{V}$. From
the Bloch-sphere perspective, this coupling acts as an effective $z$-axis
field on the time qubit whose strength and sign depend on the system
observable $\hat{V}$. Entanglement between $\mathcal{H}_{\mathrm{T}}$ and
$\mathcal{H}_{\mathrm{S}}$ then appears as a conditional rotation of the time
qubit's Bloch vector, and measurements in different Pauli bases correspond to probing
different coherent superpositions of advanced and retarded evolutions.

In this way, the Bloch-sphere representation makes the analogy to an ordinary spin qubit completely explicit: the poles represent definite temporal directions, equatorial points represent coherent superpositions of forward and backward evolution, and the length and orientation of the Bloch vector quantify the degree of coherence and the effective arrow of time.

\section{Bell inequalities}

The time qubit can be treated on the same footing as any other two-level system, which allows us to formulate Bell-type inequalities for correlations
between temporal orientation and a second qubit degree of freedom (for instance
the spin of the clock).  To make this explicit, we restrict the system Hilbert space $\mathcal{H}_{\mathrm{S}}$ to a two-dimensional subspace spanned by spin states $|\!\uparrow_z\rangle$ and $|\!\downarrow_z\rangle$, and we
consider the bipartite qubit system
\(\mathcal{H}_{\mathrm{T}}\otimes\mathcal{H}_{\mathrm{S}}\) with Pauli operators $\boldsymbol{\tau}=(\tau_x,\tau_y,\tau_z)$ acting on the time qubit
and $\boldsymbol{\sigma}=(\sigma_x,\sigma_y,\sigma_z)$ acting on the spin.

A maximally entangled state between the time qubit and the spin may be written as
\begin{equation}
    |\Phi^+\rangle_{\mathrm{TS}}
      = \frac{1}{\sqrt{2}}
        \bigl(
          |+\rangle_{\mathrm{T}}\otimes|\!\uparrow_z\rangle
        + |-\rangle_{\mathrm{T}}\otimes|\!\downarrow_z\rangle
        \bigr),
    \label{eq:TimeSpinBellState}
\end{equation}
which correlates the temporal orientation with the spin direction.  Such a
state can be prepared, for example, by starting from
$|+\rangle_{\mathrm{T}}\otimes|\!\uparrow_z\rangle$, evolving with the
controlled Hamiltonian $\hat{H}_{\mathrm{tot}}$, and post-selecting on an
appropriate spin measurement.

To formulate a Bell inequality of CHSH type, we introduce two dichotomic observables on the time qubit,
\begin{equation}
    \hat{A}_0 = \boldsymbol{a}_0\!\cdot\!\boldsymbol{\tau},
    \qquad
    \hat{A}_1 = \boldsymbol{a}_1\!\cdot\!\boldsymbol{\tau},
\end{equation}
and two dichotomic observables on the spin,
\begin{equation}
    \hat{B}_0 = \boldsymbol{b}_0\!\cdot\!\boldsymbol{\sigma},
    \qquad
    \hat{B}_1 = \boldsymbol{b}_1\!\cdot\!\boldsymbol{\sigma},
\end{equation}
where $\boldsymbol{a}_j$ and $\boldsymbol{b}_k$ are unit vectors on the
respective Bloch spheres. In this framework, $\hat{A}_0$ and $\hat{A}_1$ correspond to noncommuting observables of the time qubit, such as measurements distinguishing definite temporal orientation ($\tau_z$) and coherent time-parity superpositions ($\tau_x$), while $\hat{B}_0$ and $\hat{B}_1$ represent two different spin measurement settings.

For a given bipartite state $\rho_{\mathrm{TS}}$, the correlation function for
settings $(j,k)$ is
\begin{equation}
    E(j,k)
      = \operatorname{Tr}\!\bigl[
          \rho_{\mathrm{TS}}\,
          \hat{A}_j\otimes\hat{B}_k
        \bigr].
\end{equation}
The CHSH combination is
\begin{equation}
    S
      = E(0,0) + E(0,1) + E(1,0) - E(1,1),
    \label{eq:CHSHDef}
\end{equation}
and any local hidden-variable theory in which the outcomes of time-orientation
measurements and spin measurements are determined by pre-existing values
satisfies the Bell inequality
\begin{equation}
    |S| \leq 2.
    \label{eq:BellBound}
\end{equation}

Quantum mechanically, the correlations of the entangled state
$|\Phi^+\rangle_{\mathrm{TS}}$ can violate this bound.  Choosing measurement
directions
\begin{equation}
    \hat{A}_0 = \tau_z,\qquad
    \hat{A}_1 = \tau_x,
\end{equation}
for the time qubit and
\begin{equation}
    \hat{B}_0 = \frac{\sigma_z+\sigma_x}{\sqrt{2}},\qquad
    \hat{B}_1 = \frac{\sigma_z-\sigma_x}{\sqrt{2}},
\end{equation}
for the spin, one finds for the state
$\rho_{\mathrm{TS}} = |\Phi^+\rangle_{\mathrm{TS}}\langle\Phi^+|$ that
\begin{equation}
    S_{\mathrm{QM}}
      = \langle\Phi^+|\hat{\mathcal{B}}|\Phi^+\rangle
      = 2\sqrt{2},
    \qquad
    \hat{\mathcal{B}}
      = \hat{A}_0\otimes\hat{B}_0
       + \hat{A}_0\otimes\hat{B}_1
       + \hat{A}_1\otimes\hat{B}_0
       - \hat{A}_1\otimes\hat{B}_1,
\end{equation}
which saturates the Tsirelson bound and violates the classical inequality
\eqref{eq:BellBound}.

In the time qubit language, this violation rules out any model in which the arrow of time (encoded by $\boldsymbol{\tau}$) carries pre-assigned values that are independent of the spin context. The observables $\hat{A}_j$ probe complementary aspects of the time qubit: $\hat{A}_0=\tau_z$ tests whether the system evolves forward or backward in time, while $\hat{A}_1=\tau_x$ tests the coherence between these two directions. Their noncommutativity, 
\begin{equation}
[\hat{A}_0,\hat{A}_1]=2i\tau_y,
\label{commutator}
\end{equation}
expresses the fundamental complementarity between definite temporal direction and superposition of time orientations. A violation of \eqref{eq:BellBound} therefore implies that temporal orientation cannot be regarded as a classical variable with predetermined values; it must be treated as a quantum observable that can exist in coherent superposition and become entangled with other degrees of freedom.

In the Mach-Zehnder realization discussed below, the observables
$\hat{A}_0$ and $\hat{A}_1$ correspond to different measurements on the
path/time degree of freedom (for example, which-path detection versus
interference at the second beam splitter), while $\hat{B}_0$ and $\hat{B}_1$
correspond to different spin measurement axes.  The same CHSH structure then
applies directly to experimentally accessible correlations between time
orientation and spin, so that any observed $|S|>2$ serves as a quantitative witness of temporal coherence and of the nonclassical character of the time qubit.

\section{Mach-Zehnder realization of the time qubit}

The abstract time qubit formalism is realized in a particularly simple way by a
Mach-Zehnder interferometer with opposite magnetic fields in the two arms.
We consider a spin-$\tfrac{1}{2}$ particle entering the interferometer through a single input port.  The spatial degree of freedom is described by the path
states $|\mathrm{u}\rangle$ (upper arm) and $|\mathrm{l}\rangle$ (lower arm),
which we identify with the logical basis of the time qubit,
\begin{equation}
    |+\rangle_{\mathrm{T}} \equiv |\mathrm{u}\rangle,
    \qquad
    |-\rangle_{\mathrm{T}} \equiv |\mathrm{l}\rangle.
\end{equation}
Thus, the path degree of freedom \emph{is} the time qubit, with Pauli operators
$\tau_x,\tau_y,\tau_z$ acting on the $\{|\mathrm{u}\rangle,|\mathrm{l}\rangle\}$
subspace.

The first beam splitter prepares a coherent superposition of the two paths.
Neglecting irrelevant phases, we take
\begin{equation}
    |0\rangle \longrightarrow
    \frac{1}{\sqrt{2}}\bigl(|\mathrm{u}\rangle+|\mathrm{l}\rangle\bigr)
    = |\!+\!_x\rangle_{\mathrm{T}},
\end{equation}
so that the time qubit is initialized in the $\tau_x$ eigenstate
$|\!+\!_x\rangle_{\mathrm{T}}$.  The initial spin state $|\chi_0\rangle$ (for
instance $|\!\uparrow_x\rangle$) defines the clock degree of freedom.  The
joint input state is therefore
\begin{equation}
    |\Psi(0)\rangle
    = |\!+\!_x\rangle_{\mathrm{T}}\otimes|\chi_0\rangle.
\end{equation}

Along the upper and lower arms we place uniform magnetic fields of equal
magnitude and opposite sign, $\pm B_z$, acting only within the interferometer
region. The corresponding spin Hamiltonians are
\begin{equation}
    \hat{H}_{\mathrm{u}}
      = \hat{H}_0 + \hat{V},
    \qquad
    \hat{H}_{\mathrm{l}}
      = \hat{H}_0 - \hat{V},
\end{equation}
where $\hat{H}_0$ contains the spin-independent kinetic and potential terms
and $\hat{V}$ is the Zeeman interaction, $\hat{V} = (\hbar\Omega/2)\sigma_z$ for a field along $z$. Identifying $\hat{H}_{+}\equiv\hat{H}_{\mathrm{u}}$ and $\hat{H}_{-}\equiv\hat{H}_{\mathrm{l}}$, the total Hamiltonian in the arms is
precisely the controlled Hamiltonian,
\begin{equation}
    \hat{H}_{\mathrm{tot}}
      = |\mathrm{u}\rangle\!\langle\mathrm{u}|
         \otimes\hat{H}_{+}
      + |\mathrm{l}\rangle\!\langle\mathrm{l}|
         \otimes\hat{H}_{-}
      = \mathbb{I}_{\mathrm{T}}\!\otimes\!\hat{H}_0
        + \tau_z\!\otimes\!\hat{V},
\end{equation}
as in the general time qubit construction.  A traversal time $T$ through the
arms then yields the joint evolution
\begin{equation}
    \hat{U}_{\mathrm{arms}}(T)
      = |\mathrm{u}\rangle\!\langle\mathrm{u}|
        \otimes e^{-i\hat{H}_{+}T/\hbar}
      + |\mathrm{l}\rangle\!\langle\mathrm{l}|
        \otimes e^{-i\hat{H}_{-}T/\hbar}.
\end{equation}
Acting on the initial state, this gives
\begin{equation}
    |\Psi(T)\rangle
    = \frac{1}{\sqrt{2}}
      \Bigl(
        |\mathrm{u}\rangle\otimes e^{-i\hat{H}_{+}T/\hbar}|\chi_0\rangle
      + |\mathrm{l}\rangle\otimes e^{-i\hat{H}_{-}T/\hbar}|\chi_0\rangle
      \Bigr),
\end{equation}
which has exactly the structure of a time qubit entangled with two opposite
temporal evolutions of the spin.

The second beam splitter implements a rotation from the $\tau_z$ basis
$\{|\mathrm{u}\rangle,|\mathrm{l}\rangle\}$ to the $\tau_x$ basis associated
with the output ports $D_1$ and $D_2$. In a standard convention,
\begin{equation}
    |D_1\rangle
      = \frac{1}{\sqrt{2}}\bigl(|\mathrm{u}\rangle+|\mathrm{l}\rangle\bigr)
      = |\!+\!_x\rangle_{\mathrm{T}},
    \qquad
    |D_2\rangle
      = \frac{1}{\sqrt{2}}\bigl(|\mathrm{u}\rangle-|\mathrm{l}\rangle\bigr)
      = |\!-\!_x\rangle_{\mathrm{T}}.
\end{equation}
Applying this transformation to the path part of $|\Psi(T)\rangle$ yields
\begin{align}
    |\Psi_{\mathrm{out}}\rangle
      &= \frac{1}{2}\Bigl[
           |D_1\rangle\otimes
             \bigl(e^{-i\hat{H}_{+}T/\hbar}
                  +e^{-i\hat{H}_{-}T/\hbar}\bigr)|\chi_0\rangle
         + |D_2\rangle\otimes
             \bigl(e^{-i\hat{H}_{+}T/\hbar}
                  -e^{-i\hat{H}_{-}T/\hbar}\bigr)|\chi_0\rangle
         \Bigr] \nonumber\\
      &= |D_1\rangle\otimes\hat{U}_{\mathrm{even}}(T)|\chi_0\rangle
       + |D_2\rangle\otimes\hat{U}_{\mathrm{odd}}(T)|\chi_0\rangle,
\end{align}
where the effective operators on the spin are
\begin{align}
    \hat{U}_{\mathrm{even}}(T)
      &= \frac{1}{2}\Bigl(
           e^{-i\hat{H}_{+}T/\hbar}
          +e^{-i\hat{H}_{-}T/\hbar}
         \Bigr),\\
    \hat{U}_{\mathrm{odd}}(T)
      &= \frac{1}{2}\Bigl(
           e^{-i\hat{H}_{+}T/\hbar}
          -e^{-i\hat{H}_{-}T/\hbar}
         \Bigr),
\end{align}
in direct correspondence with Eqs.~\eqref{eq:EvenOdd}.  Detection at $D_1$
projects the time qubit onto the $+1$ eigenstate of $\tau_x$ and the spin onto
its time-even component, while detection at $D_2$ projects onto the $-1$
eigenstate and the time-odd component.  In the language of quantum operations,
$\hat{U}_{\mathrm{even}}(T)$ and $\hat{U}_{\mathrm{odd}}(T)$ are precisely the
Kraus operators associated with the two possible outcomes of a projective
measurement of $\tau_x$ on the time qubit, conditioned on its preparation in
$|\!+\!_x\rangle_{\mathrm{T}}$.

For the Zeeman Hamiltonians
$\hat{H}_{\pm} = \hat{H}_0 \pm (\hbar\Omega/2)\sigma_z$, the spin evolution in
each arm is a Larmor precession with opposite sense,
$e^{-i\hat{H}_{\pm}T/\hbar} = e^{-i\hat{H}_0 T/\hbar}
 e^{\mp i(\Omega T/2)\sigma_z}$.  Up to the common factor
$e^{-i\hat{H}_0 T/\hbar}$, the effective operators reduce to
\begin{align}
    \hat{U}_{\mathrm{even}}(T)
      &= \cos\!\left(\frac{\Omega T}{2}\right)\mathbb{I}, \\
    \hat{U}_{\mathrm{odd}}(T)
      &= -i\sin\!\left(\frac{\Omega T}{2}\right)\sigma_z.
\end{align}
Acting on an initial spin state $|\chi_0\rangle = |\!\uparrow_x\rangle$, the
detection probabilities at the two output ports are
\begin{align}
    P_{D_1}(T)
      &= \bigl\|\hat{U}_{\mathrm{even}}(T)|\!\uparrow_x\rangle\bigr\|^2
       = \cos^2\!\left(\frac{\Omega T}{2}\right),\\
    P_{D_2}(T)
      &= \bigl\|\hat{U}_{\mathrm{odd}}(T)|\!\uparrow_x\rangle\bigr\|^2
       = \sin^2\!\left(\frac{\Omega T}{2}\right).
\end{align}
These oscillations as a function of the Larmor phase $\Omega T$ are the
time-parity fringes: they arise from interference between the forward-
and backward-evolving components of the spin clock.  Which-path measurements
in the arms correspond to projective measurements of $\tau_z$ (definite
temporal orientation), while interference at the second beam splitter
implements a measurement of $\tau_x$ (time-parity). Together with spin
measurements in different bases, this setup realizes precisely the time-spin observables $\hat{A}_j$ and $\hat{B}_k$ used in the Bell analysis, making the quantum nature of the arrow of time directly accessible to
experiment.

\section{Matter and Anti-matter}

In the nonrelativistic examples discussed above, the time qubit was introduced as an auxiliary degree of freedom that controls the sign of a Hamiltonian. In
relativistic quantum theory this structure is already present in disguise: the Dirac equation admits positive- and negative-energy solutions that are mixed by motion, and these two branches can be viewed as the logical states of a time qubit. In this section we show how the Dirac Hamiltonian can be written as the coupling of a spin qubit to a time qubit, and how the operators $\tau_z$ and $\tau_x$ acquire a direct physical interpretation as, respectively, an ``energy-sign'' observable and a generator of mixing between opposite temporal orientations.

For a free relativistic particle the dispersion relation
\begin{equation}
    E^2 = p^2 c^2 + m^2 c^4
    \label{eq:RelDispersion}
\end{equation}
has two branches,
\begin{equation}
    E_\pm(\mathbf{p}) = \pm \sqrt{m^2 c^4 + c^2 \mathbf{p}^2},
\end{equation}
corresponding to opposite signs of energy. We interpret this binary degree of
freedom as a time qubit, with basis states
\begin{equation}
    |+\rangle_{\mathrm{T}} \;\leftrightarrow\; E_+(\mathbf{p}), \qquad
    |-\rangle_{\mathrm{T}} \;\leftrightarrow\; E_-(\mathbf{p}),
\end{equation}
spanning a two-dimensional Hilbert space $\mathcal{H}_{\mathrm{T}}$. The Pauli
operator $\tau_z$ then measures the sign of energy:
\begin{equation}
    \tau_z |+\rangle_{\mathrm{T}} = + |+\rangle_{\mathrm{T}},\qquad
    \tau_z |-\rangle_{\mathrm{T}} = - |-\rangle_{\mathrm{T}}.
\end{equation}
In this sense $\tau_z$ is the ``energy-sign'' operator: its eigenvalues label
forward and backward temporal orientations, that is, motion with $E>0$ and $E<0$.

The other Pauli operators $\tau_x$ and $\tau_y$ act as generators of coherent
superpositions of these two energy-sign states. For example,
\begin{equation}
    |\!\pm_x\rangle_{\mathrm{T}}
      = \frac{1}{\sqrt{2}}\bigl(|+\rangle_{\mathrm{T}}
                               \pm|-\rangle_{\mathrm{T}}\bigr),
\end{equation}
are eigenstates of $\tau_x$ with no definite energy sign: they represent
states in which forward and backward temporal orientations are maximally
superposed.

The internal Hilbert space of a relativistic spin-$\tfrac{1}{2}$ particle is
then taken to be
\begin{equation}
    \mathcal{H}
      = \mathcal{H}_{\mathrm{T}} \otimes \mathcal{H}_{\mathrm{S}},
\end{equation}
where $\mathcal{H}_{\mathrm{S}}$ is the two-dimensional spin space with Pauli
operators $\boldsymbol{\sigma}=(\sigma_x,\sigma_y,\sigma_z)$. A four-component
Dirac spinor is thus a composite state
\begin{equation}
    \psi \in \mathcal{H}_{\mathrm{T}}\!\otimes\!\mathcal{H}_{\mathrm{S}},
\end{equation}
combining a time qubit (energy sign) and a spin qubit.

The Dirac Hamiltonian must be linear in momentum so that time and space derivatives appear symmetrically in the relativistic wave equation. On the
composite space $\mathcal{H}_{\mathrm{T}}\otimes\mathcal{H}_{\mathrm{S}}$ we
therefore look for a Hamiltonian of the form
\begin{equation}
    \hat{H}_{\mathrm{D}}
      = \hat{A} \otimes m c^2
       + \hat{B}\otimes c\,\boldsymbol{\sigma}\!\cdot\!\hat{\mathbf{p}},
    \label{eq:GeneralDiracAnsatz}
\end{equation}
with $\hat{A}$ and $\hat{B}$ Hermitian operators acting on the time qubit.
Reproducing the relativistic dispersion relation \eqref{eq:RelDispersion}
requires that
\begin{equation}
    \hat{H}_{\mathrm{D}}^2
      = \mathbb{I}_{\mathrm{T}}\otimes
        \bigl[m^2 c^4 + c^2 \hat{\mathbf{p}}^{\,2}\bigr],
\end{equation}
up to the identity on internal space. Substituting
\eqref{eq:GeneralDiracAnsatz} and using
$(\boldsymbol{\sigma}\!\cdot\!\hat{\mathbf{p}})^2
 = \hat{\mathbf{p}}^{\,2}\mathbb{I}_{\mathrm{S}}$ gives
\begin{equation}
    \hat{H}_{\mathrm{D}}^2
      = \hat{A}^2\otimes m^2 c^4
       + \hat{B}^2\otimes c^2\hat{\mathbf{p}}^{\,2}
       + mc^3\,\bigl(\{\hat{A},\hat{B}\}\otimes
         \boldsymbol{\sigma}\!\cdot\!\hat{\mathbf{p}}\bigr).
\end{equation}
Thus we must have
\begin{equation}
    \hat{A}^2 = \hat{B}^2 = \mathbb{I}_{\mathrm{T}}, \qquad
    \{\hat{A},\hat{B}\} = 0.
\end{equation}
The simplest choice is
\begin{equation}
    \hat{A} = \tau_z, \qquad
    \hat{B} = \tau_x,
\end{equation}
which yields
\begin{equation}
    \hat{H}_{\mathrm{D}}
      = \tau_z \otimes m c^2
       + \tau_x \otimes c\,\boldsymbol{\sigma}\!\cdot\!\hat{\mathbf{p}}.
    \label{eq:DiracHamiltonianFromTimeQubit}
\end{equation}

This expression makes the roles of $\tau_z$ and $\tau_x$ transparent. The mass term \(\tau_z\otimes mc^2\) is diagonal in the basis $\{|+\rangle_{\mathrm{T}},|-\rangle_{\mathrm{T}}\}$.
At rest ($\mathbf{p}=0$) it assigns energies $\pm mc^2$ to these two states. Thus $\tau_z$ measures the sign of rest energy: its eigenstates are the positive- and negative-energy (matter/antimatter) components. On the time-qubit Bloch sphere this term is a static field along the $z$-axis, trying to align the Bloch vector with $\tau_z=\pm 1$. The momentum term \(\tau_x\otimes c\,\boldsymbol{\sigma}\!\cdot\!\hat{\mathbf{p}}\) is off-diagonal in the $\tau_z$ basis. It converts $|+\rangle_{\mathrm{T}}$ into $|-\rangle_{\mathrm{T}}$ and vice versa, with a strength proportional to $|\mathbf{p}|$. In other words, motion through space drives transitions between energy-sign sectors. On the Bloch sphere this is a transverse field along the $x$-axis, which causes the time-qubit Bloch vector to precess away from a definite temporal orientation.

Squaring \eqref{eq:DiracHamiltonianFromTimeQubit} gives
\begin{equation}
    \hat{H}_{\mathrm{D}}^2
      = \mathbb{I}_{\mathrm{T}} \otimes
         \bigl[m^2 c^4 + c^2 (\boldsymbol{\sigma}\!\cdot\!\hat{\mathbf{p}})^2\bigr]
      = \mathbb{I}_{\mathrm{T}}\otimes
         \bigl[m^2 c^4 + c^2 \hat{\mathbf{p}}^{\,2}\bigr],
\end{equation}
as required. The standard Dirac matrices appear as
\begin{equation}
    \beta = \tau_z \otimes \mathbb{I}_{\mathrm{S}}, \qquad
    \boldsymbol{\alpha} = \tau_x \otimes \boldsymbol{\sigma},
\end{equation}
and one may define
\begin{equation}
    \gamma^0 = \beta = \tau_z\otimes\mathbb{I}_{\mathrm{S}}, \qquad
    \gamma^i = \beta\alpha_i = i\tau_y\otimes\sigma_i,
\end{equation}
which obey $\{\gamma^\mu,\gamma^\nu\}=2\eta^{\mu\nu}\mathbb{I}$. In this
representation the Lorentz scalar
\begin{equation}
    \bar{\psi}\psi
      = \psi^\dagger\gamma^0\psi
      = \psi^\dagger(\tau_z\otimes\mathbb{I}_{\mathrm{S}})\psi
\end{equation}
is expressed directly in terms of $\tau_z$, while the probability density
remains $j^0=\psi^\dagger\psi$ in the usual way. Thus $\tau_z$ plays the role
of $\gamma^0$: it encodes the indefinite time signature of the relativistic
metric and separates positive- and negative-energy components.

The action of $\tau_z$ and $\tau_x$ is most clearly visualized on the
time-qubit Bloch sphere. Fix a momentum $\mathbf{p}$ and choose spin states
$|\chi_s(\mathbf{p})\rangle$, $s=\pm1$, such that
\begin{equation}
    \boldsymbol{\sigma}\!\cdot\!\mathbf{p}\,
    |\chi_s(\mathbf{p})\rangle
      = s|\mathbf{p}|\,
        |\chi_s(\mathbf{p})\rangle.
\end{equation}
On the two-dimensional subspace
$\mathcal{H}_{\mathrm{T}}\otimes\mathrm{span}\{|\chi_s(\mathbf{p})\rangle\}$,
the Dirac Hamiltonian reduces to an operator acting only on the time qubit,
\begin{equation}
    \hat{H}_{\mathrm{D}}^{(s)}(\mathbf{p})
      = \langle\chi_s(\mathbf{p})|
        \hat{H}_{\mathrm{D}}
        |\chi_s(\mathbf{p})\rangle
      = mc^2\,\tau_z
        + s\,c|\mathbf{p}|\,\tau_x.
\end{equation}
Writing this as
\begin{equation}
    \hat{H}_{\mathrm{D}}^{(s)}(\mathbf{p})
      = \boldsymbol{B}_{\mathrm{eff}}^{(s)}(\mathbf{p})
        \!\cdot\!\boldsymbol{\tau},
\end{equation}
we identify an effective field
\begin{equation}
    \boldsymbol{B}_{\mathrm{eff}}^{(s)}(\mathbf{p})
      = \bigl(
          s\,c|\mathbf{p}|,\,
          0,\,
          mc^2
        \bigr),
\end{equation}
with magnitude
\begin{equation}
    \bigl|\boldsymbol{B}_{\mathrm{eff}}^{(s)}(\mathbf{p})\bigr|
      = \sqrt{m^2 c^4 + c^2 \mathbf{p}^2}
      \equiv E(\mathbf{p}).
\end{equation}
In this Bloch-sphere picture, the mass term $mc^2\,\tau_z$ is a longitudinal field along the $z$-axis, favoring definite energy sign and definite temporal orientation, while the momentum term $s\,c|\mathbf{p}|\,\tau_x$ is a transverse field along $x$, which mixes the two orientations and causes the time-qubit Bloch vector to precess in the $x$-$z$ plane.

The eigenstates of $\hat{H}_{\mathrm{D}}^{(s)}(\mathbf{p})$ are time-qubit
states whose Bloch vectors are aligned or anti-aligned with
$\boldsymbol{B}_{\mathrm{eff}}^{(s)}(\mathbf{p})$, with eigenvalues
$\pm E(\mathbf{p})$. In the massless limit $m\to0$ the field lies entirely
along $x$,
\begin{equation}
    \boldsymbol{B}_{\mathrm{eff}}^{(s)}(\mathbf{p})
      \xrightarrow[m\to0]{}\,
      (s\,c|\mathbf{p}|,0,0),
\end{equation}
so the eigenstates are eigenstates of $\tau_x$ on the equator: the particle is
an equal superposition of forward and backward temporal orientations. A
nonzero mass tilts $\boldsymbol{B}_{\mathrm{eff}}^{(s)}(\mathbf{p})$ away from the equator toward $z$, energetically stabilizing a definite sign of $\tau_z$.

Time evolution under $\hat{H}_{\mathrm{D}}^{(s)}(\mathbf{p})$ corresponds to
precession of the time-qubit Bloch vector around
$\boldsymbol{B}_{\mathrm{eff}}^{(s)}(\mathbf{p})$. Superpositions of
positive- and negative-energy components therefore give oscillatory motion of
the Bloch vector between regions with $\tau_z>0$ and $\tau_z<0$. This is
precisely the internal dynamics that underlies zitterbewegung: the rapid
beating between different energy-sign components of the Dirac spinor is the
precession of the time qubit in the effective field generated by mass and
momentum.

The specific choice $\hat{A}=\tau_z$, $\hat{B}=\tau_x$ is not unique. For
example, replacing $\tau_x$ by $\tau_y$ gives
\begin{equation}
    \hat{H}' = \tau_z \otimes m c^2
             + \tau_y \otimes c\,\boldsymbol{\sigma}\!\cdot\!\hat{\mathbf{p}},
\end{equation}
which satisfies the same Clifford algebra and has the same spectrum. The two
Hamiltonians are related by a rotation of the time-qubit basis about $\tau_z$,
\begin{equation}
    \hat{U}_{\mathrm{T}} = e^{-i(\pi/4)\tau_z}, \qquad
    \hat{U}_{\mathrm{T}}\tau_x\hat{U}_{\mathrm{T}}^\dagger = \tau_y,\qquad
    \hat{H}' = \hat{U}_{\mathrm{T}}\hat{H}_{\mathrm{D}}\hat{U}_{\mathrm{T}}^\dagger.
\end{equation}
In the time-qubit language, gamma-matrix representation freedom is simply the
freedom to choose axes on the temporal Bloch sphere, without changing the
underlying physics.

Finally, the distinction between Dirac and Majorana fermions acquires a simple
geometric meaning. For a Dirac fermion, the eigenstates of $\tau_z$ (definite
energy sign and temporal orientation) can be assigned distinct conserved
charges, so particle and antiparticle correspond to opposite poles of the
time qubit Bloch sphere. A Majorana fermion, by contrast, is its own charge
conjugate: it cannot be distinguished from its antiparticle by any conserved
quantum number. In the time qubit picture this corresponds to choosing
eigenstates of a transverse operator such as $\tau_x$,
\begin{equation}
    |\!\pm_x\rangle_{\mathrm{T}}
      = \frac{1}{\sqrt{2}}\bigl(|+\rangle_{\mathrm{T}}
                               \pm|-\rangle_{\mathrm{T}}\bigr),
\end{equation}
as the fundamental modes. A Majorana mass term may be viewed as a coupling
that aligns the time qubit along such a transverse axis rather than along
$\tau_z$, pinning the physical eigenstates to coherent superpositions of
forward and backward temporal orientations instead of definite energy sign.

In summary, the Dirac Hamiltonian can be interpreted as the dynamics of a spin
qubit coupled to a time qubit. The operator $\tau_z$ measures the sign of
relativistic energy and distinguishes matter from antimatter, while $\tau_x$
generates the mixing of these two sectors induced by spatial motion. The
zitterbewegung and matter-antimatter structure of the Dirac theory thus
reflect the precession of a temporal Bloch vector in an effective field
determined by mass and momentum.

\section{Discussion}

The notion of a time qubit- a two-level system encoding opposite orientations of temporal evolution, provides an operational framework for treating the arrow of time as a genuine quantum degree of freedom. By coupling the sign of a Hamiltonian to a controllable qubit, we have shown that forward and backward evolutions can coexist coherently, interfere, and become entangled with other observables such as spin. The resulting effects, including time-parity interference and correlations between spin and temporal orientation, suggest that the direction of time can be treated on the same footing as any other internal quantum variable. In this framework, temporal direction is not imposed externally but arises as a measurable property of the system’s state.

The time qubit formulation places ideas from time-symmetric quantum theory within a concrete Hamiltonian setting. It parallels the two-state-vector formalism, in which forward- and backward-evolving components coexist, and relates to the framework of indefinite causal order, where the temporal sequence of operations may exist in superposition. By representing temporal orientation as a qubit, these concepts acquire an operational basis that can be implemented in interferometric systems. The Mach–Zehnder and spin-interferometric realizations proposed here show that standard experimental tools, such as beam splitters, magnetic fields, and phase control, are sufficient to generate and detect superpositions of opposite time orientations. In this sense, the time qubit provides a minimal and experimentally accessible model of an indefinite arrow of time.

The extension of this formalism to the relativistic regime provides a natural interpretation of matter–antimatter symmetry. The Dirac equation can be viewed as describing a spin qubit coupled to a time qubit, with the operators $\tau_z$ and $\tau_x$ representing, respectively, the sign of energy and the mixing between temporal orientations induced by spatial motion. Positive- and negative-energy solutions correspond to opposite poles of a temporal Bloch sphere, while zitterbewegung arises from precession of the time qubit Bloch vector. Within this language, pair creation and annihilation correspond to coherent flips of the time qubit, and the distinction between Dirac and Majorana fermions reflects different orientations of this temporal degree of freedom. This reinterpretation unifies the algebraic structure of relativistic quantum mechanics with the physical intuition of time symmetric evolution.

Conceptually, the time qubit framework provides a continuous bridge between reversible quantum dynamics and the classical emergence of an arrow of time. Environmental decoherence, measurement, or thermodynamic irreversibility can be understood as processes that suppress coherence between the two temporal orientations, driving the system toward a definite value of $\tau_z$. The framework thus complements current efforts in quantum thermodynamics to describe irreversibility as a dynamical, rather than postulated, feature of quantum systems.

Beyond its foundational role, the time-qubit formalism may have practical applications. Temporal superpositions could be used to construct interferometric protocols that are sensitive to Hamiltonian asymmetries or to design time-symmetric control operations in quantum information processing. Coupling between the system and a time qubit could enable conditional gate operations based on temporal orientation or provide new approaches to simulating time-reversal dynamics under controlled conditions. These ideas remain exploratory, but they suggest that time coherence could become an experimentally tunable resource.

Future work should examine how external fields and spacetime curvature influence a time qubit. Gravitational redshift and time dilation modify the relative phase between forward- and backward-evolving components, suggesting that gravity couples directly to the temporal Bloch vector. Weak gravitational gradients could thus act as effective rotations or dephasing fields on the time qubit, a possibility testable with interferometric setups sensitive to small phase shifts. Such studies may clarify how temporal coherence behaves under gravitational time dilation and could connect the time qubit picture with relativistic quantum information and the study of proper time in curved spacetime.

In summary, the time qubit framework reinterprets the direction of time as an internal quantum variable rather than a fixed background parameter. It integrates time-symmetric dynamics, relativistic matter–antimatter structure, and interferometric measurability into a single, coherent formalism. By doing so, it opens a path toward understanding temporal orientation as a manipulable aspect of quantum systems, both conceptually and, potentially, technologically.

\bibliography{citations}

\end{document}